\documentclass[runningheads]{llncs}
\usepackage{graphicx}
\graphicspath{{./figures/}}
\usepackage[utf8]{inputenc}
\usepackage{amsmath}
\usepackage{graphicx}
\usepackage{amssymb}
\usepackage{textcomp}

\newcommand{\param}{\mathbf{m}}
\newcommand{\speed}{\mathbf{v}}
\newcommand{\pos}{\mathbf{p}}
\newcommand{\normal}{\mathbf{n}}
\newcommand{\Identity}{\mathbf{I}}
\newcommand{\matSup}{\mathbf{M}}
\newcommand{\matInf}{\mathbf{N}}
\newcommand{\matB}{\mathbf{B}}
\newcommand{\matC}{\mathbf{C}}

\begin{document}

\title{Robust Cochlear Modiolar Axis Detection in CT}

\author{Wilhelm Wimmer\inst{1,2,3}\orcidID{0000-0001-5392-2074} \and
        Clair Vandersteen\inst{1,4} \and
        Nicolas Guevara\inst{1,4} \and
        Marco Caversaccio\inst{2,3,4} \and
        Herv\'{e} Delingette\inst{4}\orcidID{0000-0001-6050-5949} }


\authorrunning{W. Wimmer et al.}

\institute{Universit\'{e} C\^{o}te d'Azur, Inria, Epione, Sophia Antipolis, France\\
\email{wilhelm.wimmer@inria.fr} \and
Department of Otolaryngology, Inselspital, University of Bern, Switzerland \and
Hearing Research Laboratory, ARTORG Center, University of Bern, Switzerland \and
Universit\'{e} C\^{o}te d'Azur, Centre Hospitalier Universitaire de Nice, Institut Universitaire de la Face et du Cou, Nice, France}

\maketitle              

\begin{abstract}
The cochlea, the auditory part of the inner ear, is a spiral-shaped organ with large morphological variability. 
An individualized assessment of its shape is essential for clinical applications related to tonotopy and cochlear implantation. 
To unambiguously reference morphological parameters, reliable recognition of the cochlear modiolar axis in computed tomography (CT) images is required. 
The conventional method introduces measurement uncertainties, as it is based on manually selected and difficult to identify landmarks.
Herein, we present an algorithm for robust modiolar axis detection in clinical CT images. We define the modiolar axis as the rotation component of the kinematic spiral motion inherent in the cochlear shape.
For surface fitting, we use a compact shape representation in a 7-dimensional kinematic parameter space based on extended Plücker coordinates. 
It is the first time such a kinematic representation is used for shape analysis in medical images. 
Robust surface fitting is achieved with an adapted approximate maximum likelihood method assuming a Student-t distribution, enabling axis detection even in partially available surface data. 
We verify the algorithm performance on a synthetic data set with cochlear surface subsets. 
In addition, we perform an experimental study with four experts in 23 human cochlea CT data sets to compare the automated detection with the manually found axes. Axes found from co-registered high resolution \textmu CT scans are used for reference.
Our experiments show that the algorithm reduces the alignment error providing more reliable modiolar axis detection for clinical and research applications.

\end{abstract}

\section{Introduction}
The cochlea is a spiral structure in the inner ear that transduces acoustic waves into electrical nerve impulses to enable hearing. The morphology of the human cochlea is complex and highly variable. Therefore, an unambiguous description of the morphological parameters for both modeling and clinical applications is required. Of great importance is the modiolar axis, the central axis of the spiral shape, as it is used to define the z-axis of cylindrical cochlear coordinate reference systems~\cite{ref_verbist2010}. Due to the tonotopic organization of the cochlea, the modiolar axis connects anatomical features with physiological parameters, mapping spatial positions along the spiral with the perceived characteristic frequencies. In cochlear implantation, in which an electrode array is inserted into the cochlea to restore hearing, radiological parameters referenced by the modiolar axis are used for preoperative planning (selection of suitable implant lengths), for postoperative evaluation (array insertion depth assessment) and for audio processor programming (patient-specific tonotopic stimulation maps). They are also used to investigate the effects of tonotopic mismatch between different stimulation channels on speech rehabilitation in bilateral cochlear implant users. 

A common definition of the modiolar axis is based on anatomical landmarks that can only be identified imprecisely in computed-tomography (CT) images: the helicotrema and the center of the modiolus in the basal turn of the cochlea~\cite{ref_wimmer2014}. This leads to misalignment and inter-observer variability, even when using multi-planar reconstructions. As a consequence, outcome measures that are referenced by modiolus-based coordinate systems are distorted. Furthermore, misclassification of cochlear morphology can be caused by inaccurate modiolar axis estimation~\cite{ref_demarcy2017}. Previous detection algorithms use center-line based methods, however either requiring fully segmented image data of the cochlea~\cite{ref_demarcy2017} or knowledge of additional extrinsic parameters~\cite{ref_yoo2000}.

Herein, we present a novel approach for modiolar axis detection suitable for clinical resolution CT images. The spiral shape of the cochlea is modelled as a kinematic surface to mimic its natural growth. Kinematic surfaces are defined as the location of surface points that are tangent to a parameterized stationary velocity field. Then, the modiolar axis is determined as the rotation component of the intrinsic spiral motion. Our contribution lies in the first application of a compact seven-dimensional kinematic surface representation for medical image analysis. Furthermore, we extend the method by a robust maximum likelihood scheme based on a Student-t distribution. The algorithm output is verified using a synthetic data set. Finally, we perform an experimental validation study to compare the modiolar axis detection results between the conventional landmark-based method and our algorithm under consideration of \textmu CT reference data.

\section{Methods}
\subsection{Kinematic Modiolar Axis Detection}
To find the modiolar axis, we aim to determine the rotation component of the intrinsic kinematic spiral motion forming the cochlea. We base our work on a line element geometry approach for kinematic surface recognition~\cite{ref_hofer2005}. A kinematic surface is a surface consisting of oriented points (with position $\pos_i$ and unit surface normals $\normal_i$) that is tangent to a parametric velocity field $\speed(\pos)$, i.e. $\speed(\pos) \cdot \normal = 0$. Kinematic surfaces somewhat extend the notion of implicit surfaces $\mathrm{S}(\pos)= 0$ defined on points. 

\begin{figure}
\centering
\includegraphics[trim=120 45 100 120, clip,width=.7\textwidth]{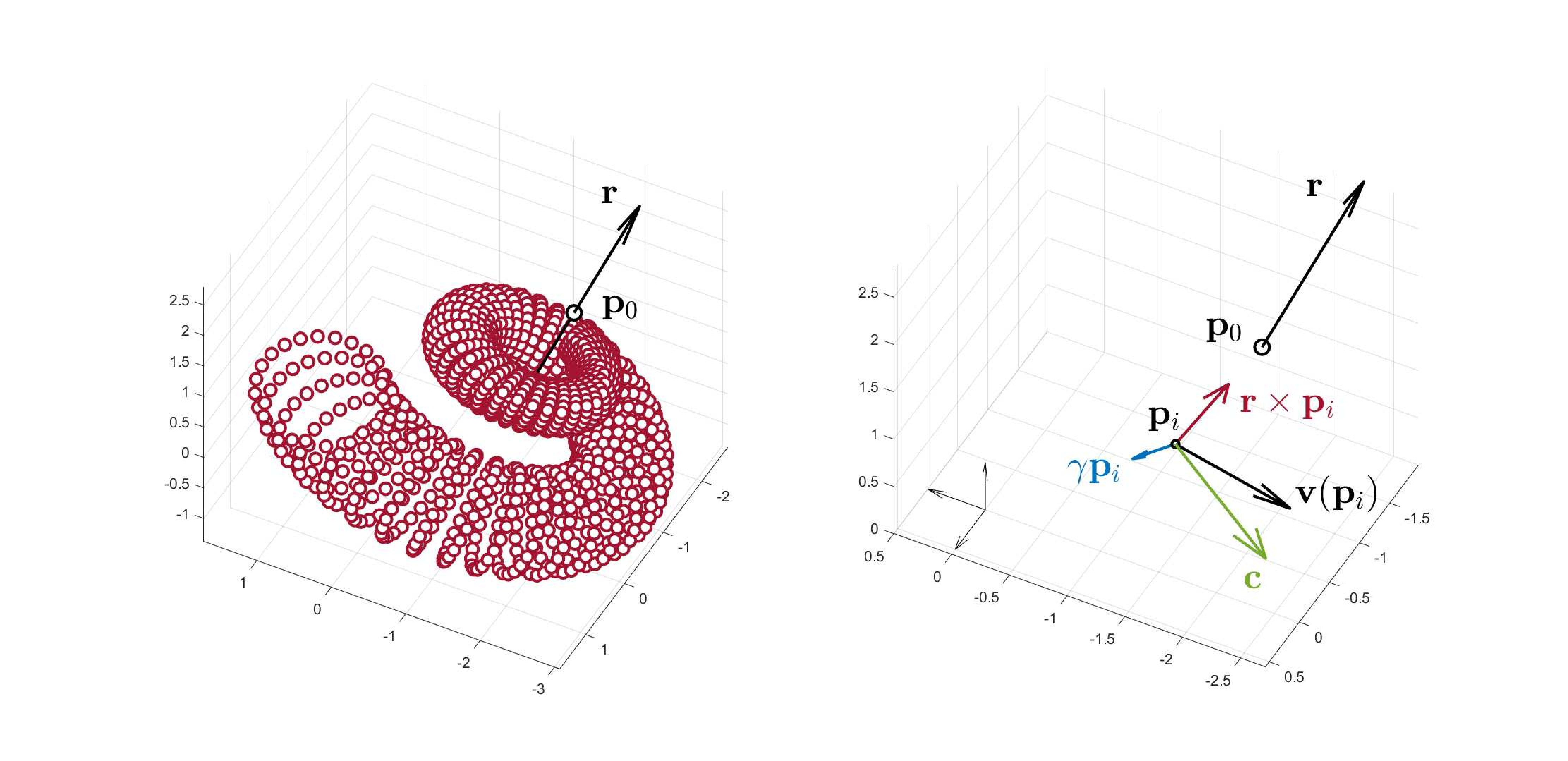}
\caption{\small{(Left) Spiral-shaped surface generated by a kinematic motion with rotation axis $\mathbf{r}$ and zero velocity convergence point $\mathbf{p}_0$. (Right) Components of the velocity $\mathbf{v}(\mathbf{p}_i)$ at point $\mathbf{p}_i$ with rotation $\mathbf{r}$, translation $\mathbf{c}$, and scaling factor $\gamma$.}} \label{fig1}
\end{figure}

For the cochlea, we consider a spiral velocity field $\speed(\pos)=\mathbf{r}\times\pos+\mathbf{c}+\gamma\pos$, consisting of a rotation $\mathbf{r}$, translation $\mathbf{c}$ and scale factor $\gamma$ (see Fig. \ref{fig1}). By choosing $\mathbf{f}(\pos,\normal) = \{ \pos \times \normal, \normal, \pos \cdot \normal \}$ the surface is transformed into a seven-dimensional parameter space based on extended Plücker coordinates. Then, the problem of spiral shape recognition reduces to fitting a linear subspace to $\mathbf{f}(\pos,\normal)$~\cite{ref_hofer2005}. This is achieved by searching for the parameters $\param=\{\mathbf{r},\mathbf{c},\gamma\}$ such that the distance $d_i(\param)$ between each point and the surface tangent to the velocity field $\speed(\pos_i)$ is minimized. By using a first order approximation of the distance (approximate maximum likelihood method~\cite{ref_andrews2013,ref_chernov2007}) we can write:

\begin{equation}
    d_i(\param)=\frac{\speed(\pos_i)\cdot\normal_i}{\sqrt{\|\speed(\pos_i)\|^2+w_\mathbf{p}\|\nabla_\mathbf{p}(\speed(\pos_i)\cdot\normal_i)\|^2}}
\end{equation}
where $w_p$ is a scalar regularizing the denominator. For a spiral velocity field, we find $\nabla_\mathbf{p}(\speed(\pos_i)\cdot\normal_i)=\normal_i \times \mathbf{r}+\gamma\normal_i=[\mathbf{A}_{\mathbf{r}}+\gamma\Identity]\normal_i$ where $\mathbf{A_{r}}$ is the skew-symmetric matrix associated with vector $\mathbf{r}$, and $\Identity$ is the identity matrix. Then we have $\|\nabla_\mathbf{p}(\speed(\pos_i)\cdot\normal_i)\|^2=\|\normal_i \times \mathbf{r}\|^2+\gamma^2$ and $\|\speed(\pos_i)\|^2=\|\mathbf{r}\times\pos_i\|^2 +\gamma^2\|\pos_i\|^2   + 2 \gamma (\pos_i\cdot\mathbf{c})+2 [\mathbf{r},\pos_i,\mathbf{c}] + \|\mathbf{c}\|^2$. In a first approach, we assume a Gaussian distribution of the distance error with variance $\Sigma$, i.e., $p(d_i)={\cal N}(d|0,\Sigma)$. The objective now is to determine $\param$ such that the log-likelihood $p(\mathcal{D}|\param)$ is maximized:
     \[ 
    \log p(\mathcal{D}|\param)=\log \prod_{i=1}^n p(d_i|\param)= -\frac{n}{2}\log{2\pi \Sigma} - \frac{1}{2} \sum_{i=1}^n \frac{d_i^2}{\Sigma} = -\frac{n}{2}\log{2\pi \Sigma} + \mathcal{L}(\param).
     \]
We can write $d_i^2=\frac{(\speed(\pos_i)\cdot\normal_i)^2}{\|\speed(\pos_i)\|^2+w_p\|\nabla_p(\speed(\pos_i)\cdot\normal_i)\|^2}=\frac{\param^T\matSup_i\param}{\param^T\matInf_i\param}$ with the $7\times 7$ matrices $\matSup_i$, $\matInf_i$ defined as:
\[
   \matSup_i=\mathbf{f}(\pos_i,\normal_i) \mathbf{f}(\pos_i,\normal_i)^T,   \quad \matInf_i=\begin{bmatrix} \mathbf{A}_{\pos_i}^T\mathbf{A}_{\pos_i} + w_p \mathbf{A}_{\normal_i}^T\mathbf{A}_{\normal_i} & -\mathbf{A}_{\pos_i} & \mathbf{0}\\ 
   -\mathbf{A}^T_{\pos_i} & \Identity & \pos_i \\ 
   \mathbf{0} & \pos_i^T & \pos_i\cdot\pos_i +  w_p \end{bmatrix}. 
\]
Maximizing the log likelihood $\mathcal{L}(\param)$ is equivalent to minimizing $\sum_{i=1}^n \frac{\param^T\matSup_i\param}{\param^T\matInf_i\param}$ which leads to solving the  generalized eigenvalue problem~\cite{ref_chernov2007}: 
\begin{equation}
\matB_m \param=\matC_m \param
\label{eq_geneig}
\end{equation} with $\matB_m=\sum_{i=1}^n \frac{\matSup_i}{\param^T\matInf_i\param}$ and $\matC_m=\sum_{i=1}^n \frac{\param^T\matSup_i\param}{(\param^T\matInf_i\param)^2}\cdot\matInf_i$.
\\

The non-linear problem can be tackled by iteratively computing the matrices $\matB_m$ and $\matC_m$ for a given estimation of $\param$ and then estimating $\param$ as the eigenvector associated with the smallest eigenvalue (closest to zero). We then define the direction of the modiolar axis as the rotation component $\mathbf{r}$ of the estimated parameters. We further need to compute the zero velocity center of the spiral motion $\mathbf{p}_0 = \frac{1}{\gamma ( \mathbf{r}^{2} + \gamma^{2})} (\gamma \mathbf{r} \times \mathbf{c} -\gamma^{2}\mathbf{c}-(\mathbf{r} \cdot \mathbf{c})\mathbf{r})$ to define the position of the modiolar axis.

\subsection{Robust Detection}
Like any least-squares fitting method, the above mentioned approach is sensitive to outliers. To increase robustness, we propose to replace the Gaussian likelihood with a Student-t distribution, which is a Gaussian Scale Mixture. More precisely, we assume $p(d_i)={\mathrm{St}}(d_i|0,\Sigma,\nu)=\int_{z_i}\mathcal{N}(d_i|0,\Sigma/z_i)\,\mathrm{Ga}(z_i|\nu/2,\nu/2)$ where $z_i$ is the variance scale variable which has a prior given by the Gamma distribution parameterized by the degrees of freedom $\nu$.  When $\nu\Longrightarrow +\infty$, then the Student-t is equivalent to the Gaussian distribution. Therefore, the $\nu$ variable is inversely proportional to the number of outliers. The estimation of the kinematic surface is now extended with an Expectation-Maximization scheme, where $z_i$ is the latent variable~\cite{ref_scheffler2008}. This is equivalent to iteratively estimating $\nu$, $z_i$ and $\Sigma$ with the following steps:
\begin{itemize}
\item{\textbf{E-step:}} Estimate $z_i$ for each data point as $z_i=(\nu+1) /(\nu+d_i^2/\Sigma)$. When $\nu$ is very large then $z_i$ is close to 1, irrespective to the Mahalanobis distance $d_i^2/\Sigma$. When $\nu$ is less large then $z_i$ is close to zero for outliers (since the Mahalanobis becomes large) and close to 1 for inliers.
\item{\textbf{M-$\Sigma$ step:} Estimate the variance as $\Sigma=\frac{1}{n} \sum_{i=1}^n z_i d_i^2$}
\item{\textbf{M-$\nu$ step:} Estimate $\nu$  as the solution of the non-linear problem
     \[ 
     -\psi\Big(\frac{\nu}{2}\Big) + \log\Big(\frac{\nu}{2}\Big) + 1 +\psi\Big(\frac{\nu+1}{2}\Big) -\log\Big(\frac{\nu+1}{2}\Big)+\frac{1}{n} \sum_{i=1}^n (\log z_i -z_i)=0,
     \] where $\psi(x)$ denotes the digamma function.}
\end{itemize}
As an output, we get the estimation of confidence $z_i$ in each data point. To obtain a robust estimation of the parameter vector $\param$, we proceed as before (\ref{eq_geneig}) but with $\matB_m=\sum_{i=1}^n z_i \frac{\matSup_i}{\param^T\matInf_i\param}$ and $\matC_m=\sum_{i=1}^n z_i \frac{\param^T\matSup_i\param}{(\param^T\matInf_i\param)^2}\matInf_i$.

\subsection{Implementation}
\begin{figure}
\centering
\includegraphics[width=0.6\textwidth]{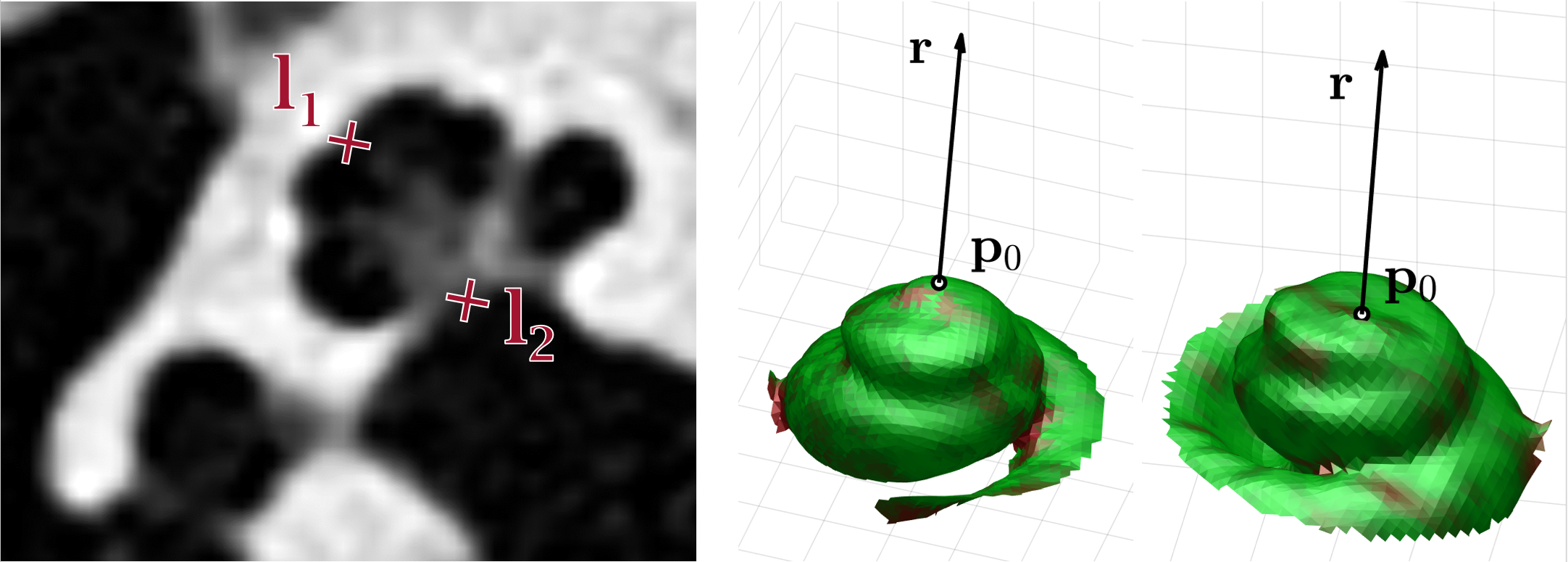}
\caption{\small{(Left) CT slice with cochlear cross-section and 2 modiolar axis landmarks: the helicotrema ($\mathrm{l_1}$) and the center of the modiolus in the basal turn ($\mathrm{l_2}$). (Right) Examples of robust fitting in partial cochlea surfaces for modiolar axis $\mathbf{r}$ and zero velocity center $\mathbf{p}_0$ detection. The confidence $z_i$ of each oriented point is color-encoded from red (close to 0, outliers) to green (close to 1, high confidence).}} \label{fig2}
\end{figure}

The detection algorithm was implemented in Matlab. It is initialized by specifying the landmarks $\mathbf{l}_1$ and $\mathbf{l}_2$ (Fig. \ref{fig2} left) in CT to preselect a spherical segment volume approximately covering the middle and apical turns of the cochlea (radius $r_{s} = \|\mathbf{r}_s\| = \|\mathbf{l}_1-\mathbf{l}_2\|$ and center $\mathbf{c}_{s} = (\mathbf{l}_1 +\mathbf{l}_2)/2$, cropped by the planes perpendicular to $\mathbf{r}_s$ in $\mathbf{l}_1$ and $\mathbf{l}_2$). This volume is chosen to minimize interference of proximal structures that may appear connected to the cochlea in clinical CT (i.e., the tympanic cavity at the round window, the internal auditory canal, and the facial nerve). The data is labelled through intensity thresholding (isovalue at 1000 HU), smoothed, and isosurfaces are generated by a marching cubes routine. The largest connected surface with the center of gravity closest to $\mathbf{c}_{s}$ is extracted. The point cloud is scaled and centered, surface normals are computed and the parameter space $\mathbf{f}(\mathbf{p})$ is obtained. Kinematic surface fitting is performed by 5 iterations with $w_p = 0.001$ (Fig. \ref{fig2} right).

\subsection{Verification in Synthetic Data}
We used a polynomial cochlea model with known modiolar axis to generate a synthetic data set for algorithm verification~\cite{ref_pietsch2017}. To mimic the facial nerve located close to the cochlea and often causing segmentation artifacts, we added a tubular structure (1 mm in diameter) with constrained random alignment. In addition, point positions were perturbed with Gaussian noise (0.15 mm standard deviation). We tested the robustness of the algorithm in varying levels of surface coverage (analogous to algorithm implementation, however with different radii covering 5\% to 100\% of total points). For each level, 500 random cochleae were generated. For comparison, the non-robust (Gaussian) version of the detection algorithm as well as simple detection using principal component analysis (PCA) of the point cloud was applied, where the modiolar axis was selected as the component with least variance. We assessed the alignment between the estimated axis $\mathbf{r}$ and reference axis $\mathbf{r}_\mathrm{ref}$ with the angular error $\Delta \theta$ and the distance error $\Delta d$ defined as the closest absolute distance between the estimated axis $\mathbf{r}$ and the center point $\mathbf{p}_\mathrm{ref}$ on the reference axis $\mathbf{r}_\mathrm{ref}$:

\begin{equation}
\Delta \theta = \arcsin{ \frac{\|\mathbf{r}_\mathrm{ref}\times \mathbf{r}  \|}{\|\mathbf{r}_\mathrm{ref} \| \|\mathbf{r}  \|}}, \qquad
\Delta d = \Bigl| (\mathbf{p}_0 - \mathbf{p}_\mathrm{ref} ) - \frac{\mathbf{r} \cdot(\mathbf{p}_0 - \mathbf{p}_\mathrm{ref} ) }{\|\mathbf{r}\|} \Bigr|
\label{eq_error}
\end{equation}

\begin{figure}
\centering
\includegraphics[width=1\textwidth]{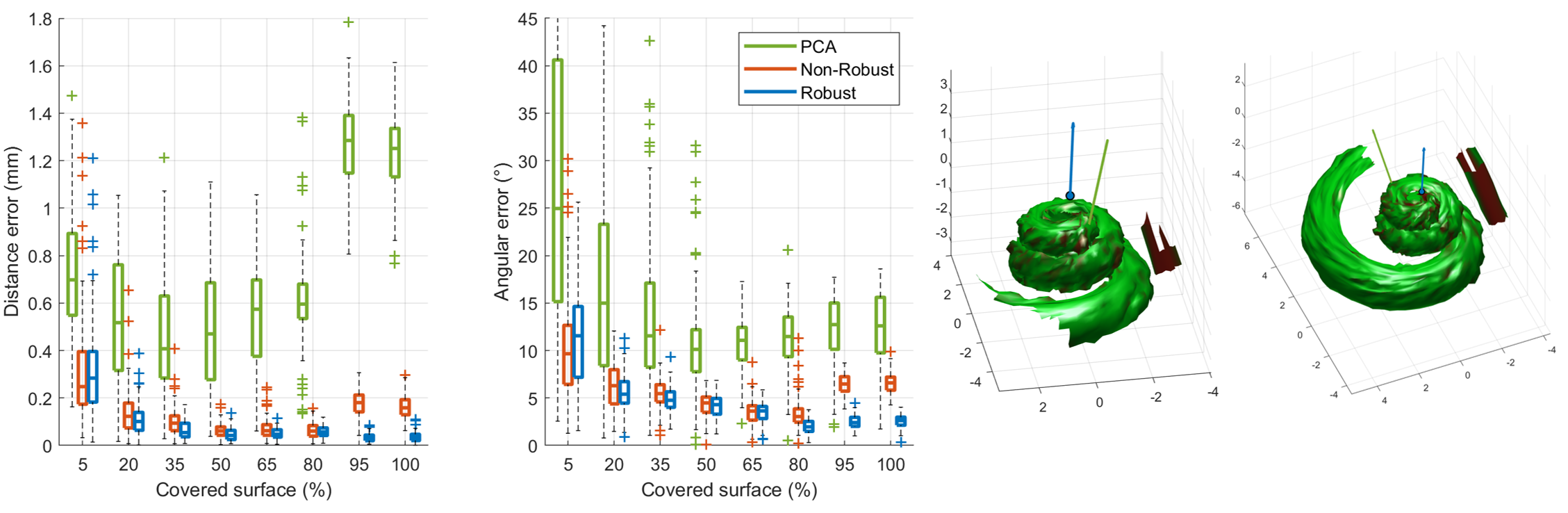}
\caption{\small{(Left) Distance and angular errors after modiolar axis detection in synthetic data for varying levels of cochlea surface coverage (percentage of extracted points vs. total number of points). The coverage of the implemented algorithm is $\sim$80\%. (Right) Two examples of robust fitting in partial cochlea surfaces with facial nerve. The confidence $z_i$ of each oriented point is color-encoded from red (close to 0, outliers) to green (close to 1, high confidence). }} \label{fig3}
\end{figure}

\subsection{Experimental Validation}
We validated the algorithm with a data set of 23 human temporal bone specimens consisting of clinical CT (voxel size: 156 $\times$ 156  $\times$ 200 \textmu m$^3$) and co-registered \textmu CT (voxel size: 60$^3$ \textmu m$^3$) scans. Four experts manually identified 2 landmarks (see Fig. \ref{fig2}) for each sample in multi-planar CT reconstructions to specify the modiolar axis. The same landmarks were also used to initialize the robust detection algorithm. Again, we applied the simple PCA-based and the non-robust axis estimations for comparison. As reference, the modiolar axis and its center were determined in each specimen using high-resolution surface models from the segmented \textmu CT data. The alignment differences were assessed with the equations shown in (\ref{eq_error}). Differences in alignment errors were estimated using (separate) linear mixed-effects models (R environment with lme4 package)~\cite{ref_lme4}, with a fixed effect for the detection method (categorical variable). We included random intercepts for specimens and experts to account for paired measurements. Before analysis, the data was log transformed.

\section{Results}
Figure \ref{fig3} illustrates the verification results. With increasing surface data available, the robust algorithm converges to an angular error of 2.5 degrees and an distance error below 0.1 mm. In contrast, the vertex-based PCA detection shows limited improvement and even yields worse distance errors with increasing cochlea surface coverage, since the basal turn vertices cause a shifting away from the modiolar axis. As expected, the non-robust (Gaussian) version of the algorithm is more sensitive to outliers (facial nerve structure).

\begin{figure}
\centering\includegraphics[trim=70 0 70 15, clip,width=0.7\textwidth]{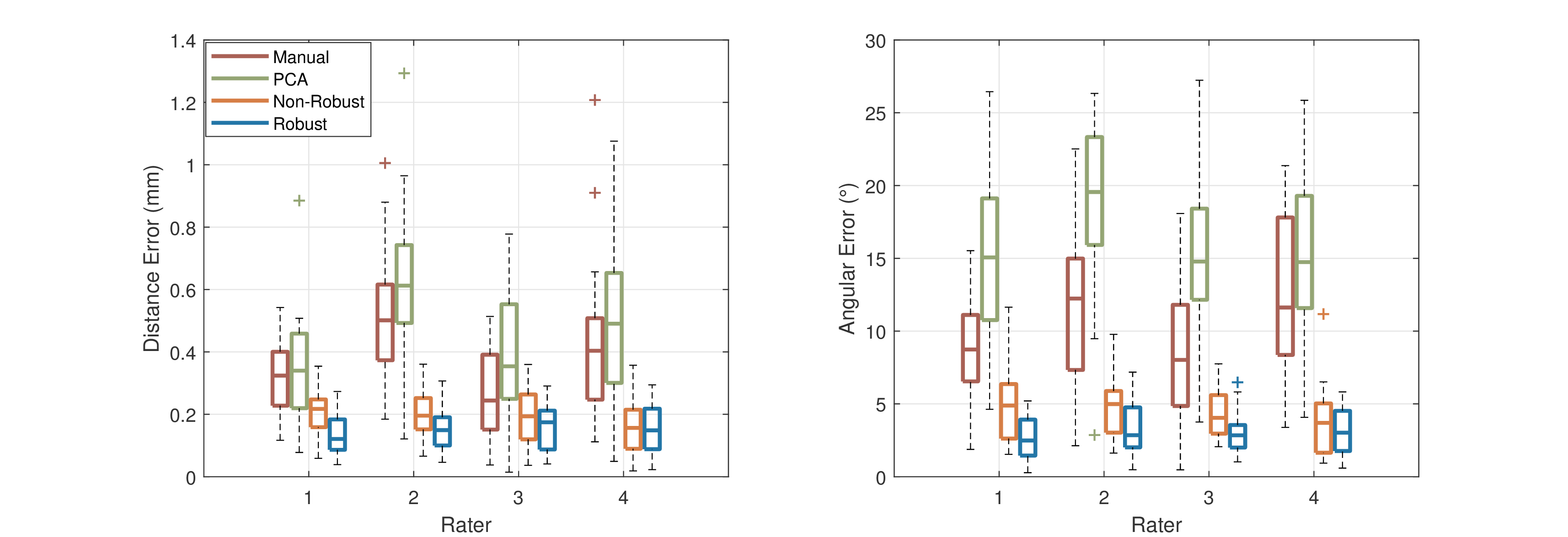}
\caption{\small{Alignment errors using manual landmark-based, PCA-based, non-robust and robust kinematic modiolar axis detection in 23 specimens.}} \label{fig4}
\end{figure}

\begin{figure}
\centering
\includegraphics[trim=120 120 90 120, clip,width=0.7\textwidth]{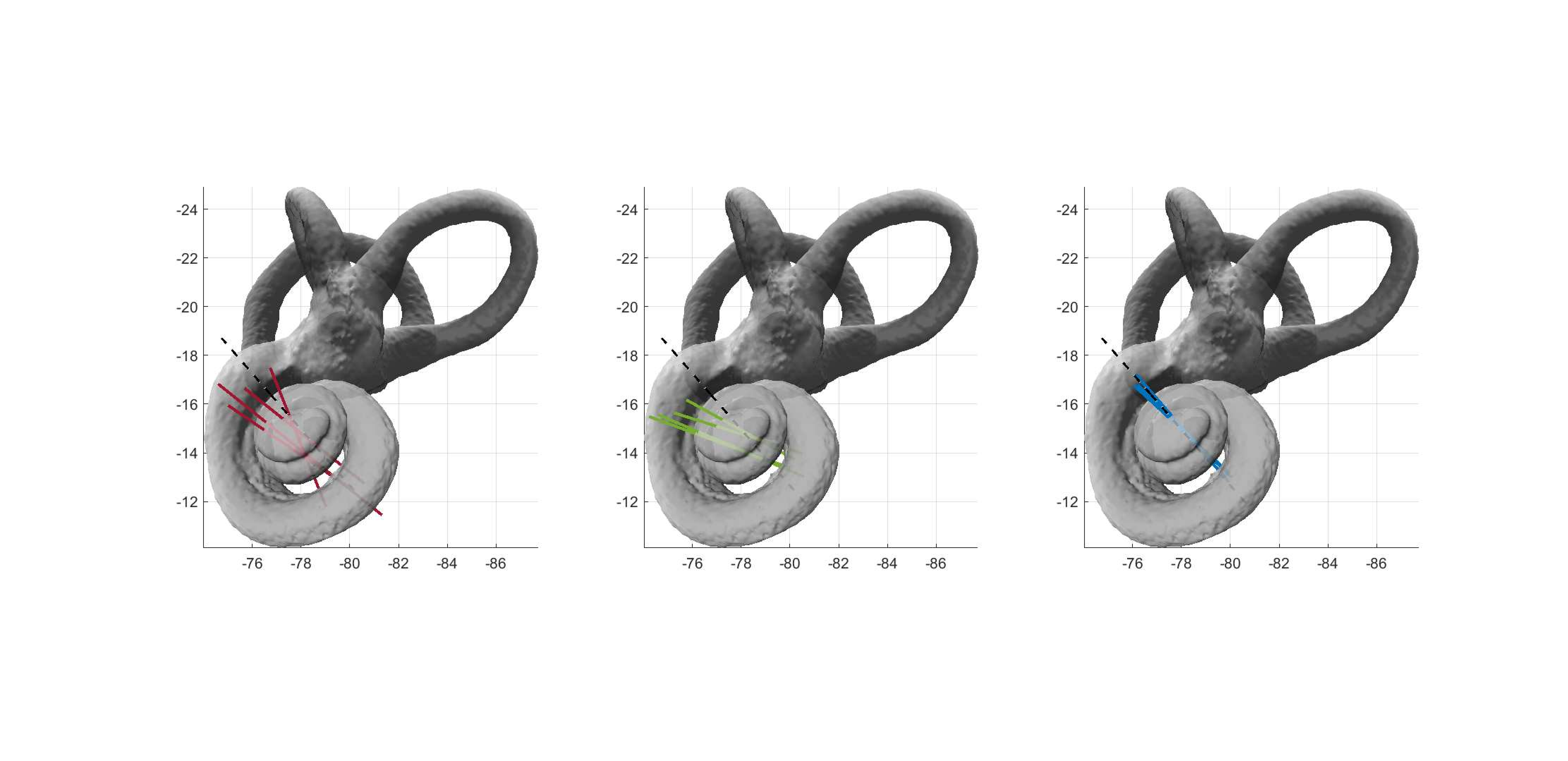}
\caption{\small{Bony labyrinth visualization (\textmu CT, specimen 11) with reference modiolar axis (dashed line). Modiolar axes after manual landmark-based (left), PCA-based (middle), and robust kinematic detection (right) in CT data are shown for comparison.} } 
\label{fig5}
\end{figure}

Figure \ref{fig4} summarizes the alignment error after manual landmark-based and automated modiolar axis detection in the CT data of the 23 specimens. The PCA based detection performs worse than manual selection. As measured by the linear mixed effects models, compared with the manual procedure, the robust method reduced the average distance error from 0.32 mm to 0.13 mm (improvement by 0.19 mm, 95\% confidence interval [0.17 mm, 0.21 mm]) and the angular error from 9.0° to 2.4° (improvement by 6.6°, 95\% confidence interval, [6.1°, 6.9°]). The non-robust version performed worse than the robust version (average distance error 0.04 mm higher and angular error 1.5° higher). The robust procedure further reduced the variability between the observers. Figure \ref{fig5} visualizes an example.

\section{Conclusions}
We present a novel and anatomically meaningful approach to model the spiral shape of the cochlea as a structure formed by a kinematic motion (natural growth) and extract the modiolar axis as the rotation component. 
This formalism enables us to detect the modiolar axis even in subsets of cochlea surfaces obtained from CT images. 
The approach generalizes the implicit surface representation of oriented points. 
The parameters $\param=\{\mathbf{r},\mathbf{c},\gamma\}$ provide a compact representation of the cochlea based on intrinsic shape properties and enable novel approaches to cochlear morphology classification. Using appropriate curves, it could be used to generate surfaces for cochlea segmentation.
The approach is further applicable for kinematic surface detection of cylindrical, conical, rotational and helical motions~\cite{ref_hofer2005}.
It could be extended by improved surface extraction methods or by directly using image gradients from CT for kinematic parameter space computation. The script runs in $\sim$3 sec on a standard laptop (Intel i7).

\subsubsection{Acknowledgments.}
Supported by the Swiss National Science Foundation (grant no. P400P2\_180822) and the French government (UCA$^{\mathrm{JEDI}}$ -  ANR-15-IDEX-01).

\bibliographystyle{splncs04}

\begin{thebibliography}{10}

\bibitem{ref_verbist2010}
Verbist, B.M., et~al.: Consensus panel on a cochlear coordinate system applicable in histologic, physiologic, and radiologic studies of the human cochlea. Otology Neurotology \textbf{31}(5),722--730 (2010)

\bibitem{ref_wimmer2014}
Wimmer, W., et~al.: Semiautomatic cochleostomy target and insertion trajectory planning for minimally invasive cochlear implantation. Biomedical Research International (2014). \doi{10.1155/2014/596498}

\bibitem{ref_demarcy2017}
Demarcy, T., et~al.: Automated analysis of human cochlea shape variability from seg-mented \textmu CT images. Computerized Medical Imaging and Graphics \textbf{59},1--12 (2017) 

\bibitem{ref_yoo2000}
Yoo, S. K., Wang, G., Rubinstein, J. T., Vannier, M. W.: Three-dimensional geometric modeling of the cochlea using helico-spiral approximation. IEEE Transactions on Biomedical Engineering, \textbf{47}(10),1392--1402 (2000)

\bibitem{ref_hofer2005}
Hofer, M., Odehnal, B., Pottmann, H., Steiner, T., Wallner, J.: 3D shape recognition and reconstruction based on line element geometry. In: 10th International Conference on Computer Vision, pp. 1532--1538. IEEE, Beijing (2005)

\bibitem{ref_andrews2013}
Andrews, J., Séquin, C.H.: Generalized, basis-independent kinematic surface fitting. Computer-Aided Design \textbf{45}(3),615--620 (2013)

\bibitem{ref_chernov2007}
Chernov, N.: On the Convergence of Fitting Algorithms in Computer Vision. Journal of Mathematical Imaging and Vision \textbf{27},231--239 (2007)  

\bibitem{ref_scheffler2008}
Scheffler, C.: A derivation of the EM updates for finding the maximum likelihood parameter estimates of the Student-t distribution. \url{http://www.inference.org.uk/cs482/publications/scheffler2008derivation.pdf}. Last accessed 29 March 2019

\bibitem{ref_pietsch2017}
Pietsch, M., et al.: Spiral Form of the Human Cochlea Results from Spatial Constraints. Scientific Reports \textbf{7}:7500 (2017)

\bibitem{ref_lme4}
Bates, D., M\"achler, M., Bolker, B., Walker, S.: Fitting Linear Mixed-Effects Models Using {lme4}. Journal of Statistical Software \textbf{67}(1),1--48 (2015)
    
\end{thebibliography}

\end{document}